\documentclass[twocolumn,showpacs,aps,pra]{revtex4-1}
\usepackage{amssymb}
\usepackage{amsfonts}
\usepackage{dcolumn}
\usepackage{amsmath}
\usepackage{graphicx}
\usepackage{psfrag,pstricks}
\usepackage{amsmath}
\usepackage{amssymb}

\begin{document}
\title{Strong squeezing and robust entanglement in cavity electromechanics }
\author{Eyob A. Sete$^1$ and Hichem Eleuch$^2$}
\affiliation{$^1$Department of Electrical Engineering, University of California, Riverside, California 92521, USA \\
$^2$Department of Physics, McGill University, Montreal, Canada H3A 2T8}
\date{\today}

\begin{abstract}
We investigate nonlinear effects in an electromechanical system consisting of a superconducting charge qubit coupled to transmission line resonator and a nanomechanical oscillator, which in turn is coupled to another transmission line resonator. The nonlinearities induced by the superconducting qubit and the optomechanical coupling play an important role in creating optomechanical entanglement as well as the squeezing of the transmitted microwave field. We show that strong squeezing of the microwave field and robust optomechanical entanglement can be achieved in the presence of moderate thermal decoherence of the mechanical mode. We also discuss the effect of the coupling of the superconducting qubit to the nanomechanical oscillator on the bistability behaviour of the mean photon number.
\end{abstract}


\maketitle

\section{Introduction}
Cavity optomechanics, where the electromagnetic mode of the cavity is coupled to the mechanical motion via radiation pressure force, has attracted a great deal of renewed interest in recent years \cite{Nori13}. Such coupling of macroscopic objects with the cavity field can be used to directly investigate the limitation of the quantum-based measurements and quantum information protocols \cite{Bra92,Man97,Bos97}. Furthermore, optomechanical coupling is a promising approach to create and manipulate quantum states of macroscopic systems. Many quantum and nonlinear effects have been theoretically investigated. Examples include, squeezing of the transmitted field \cite{Fab94,Woo08,Set12}, entanglement between the cavity mode and the mechanical oscillator \cite{Zha03,Pin05,Vit08}, optical bistability \cite{Tre96,Dor83,Mey85,Jia12,Set12}, side band ground state cooling \cite{Oco10,Teu11} among others. In particular, the squeezing of the transmitted field and the optomechanical entanglement strongly rely on the nonlinearity induced by the optomechanical interaction which couples the position of the oscillator to the intensity of the cavity mode. Recently, relatively strong optomechanical squeezing has been realized experimentally by exploiting the quantum nature of the mechanical interaction between the cavity mode and a membrane mechanical oscillator embedded in an optical cavity \cite{Pur13}.

On the other hand, demonstrations of ground state cooling, manipulation, and detection of mechanical states at the quantum level require strong coupling, where the rate of energy exchange between the mechanical oscillator and the cavity field exceeds the rates of dissipation of energy from either system. Although the control and measurement of a single microwave phonon has already been demonstrated \cite{Oco10}, the phonon states appeared to be short-lived. However, for practical applications mechanical states should survive longer than the operation time. This unwanted property is due to the fact that mechanical resonators performance degrades as the fundamental frequency increases \cite{Eki05}.

In order to observe the quantum mechanical effects in cavity optomechanics, one needs to reach the strong coupling regime and overcome the thermal decoherence. This has been exceedingly difficult to experimentally demonstrate in cavity optomechanics schemes. An alternative approach to realize strong coupling is to use electromechanical systems, where the mechanical motion is coupled to superconducting circuits embedded in transmission line resonators \cite{Rab04,Gel04,Gel05,Zho06,Wen08,Wen082,Teu11,Nic12,Yin12}. Teuful \textit{et al}. \cite{Teu11} have recently realized strong coupling and quantum enabled regimes using electromechanical systems composed of low-loss superconducting circuits. These systems fulfil the requirements for experimentally observing and controlling the theoretically predicted quantum effects \cite{Zha03,Pin05,Vit08,Tre96,Dor83,Mey85,Jia12,Set12}. In this regard, much attention has been paid in exploiting experimentally accessible electromechanical systems \cite{Rab04,Gel04,Gel05,Zho06,Wen08,Wen082,Nic12,Yin12}.

\begin{figure}[t]
\includegraphics[width=5cm]{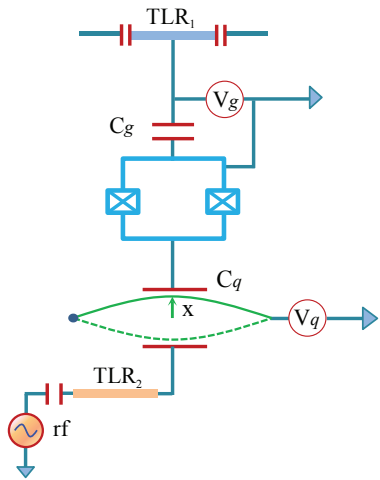}
\caption{Schematics of our model. A Cooper pair box, consists of two Josephson junctions, is coupled to a superconducting transmission line resonator ($\text{TLR}_{1}$) and a nanomechanical oscillator. In general, the interaction between the qubit (the Cooper pair box) and the nanomechanical oscillator is nonlinear, which depends on the variable capacitor, $C_{q}$. A second superconducting transmission line resonator ($\text{TLR}_{2}$) is capacitively coupled to the nanomechanical oscillator. The radio frequency (rf) source produces a microwave field, which populates the second resonator $\text{TLR}_{2}$ via a small capacitance. }\label{fig1}
\end{figure}

In this work, we investigate the squeezing and the optomechanical entanglement, in an electromechanical system in which a superconducting charge qubit is coupled to a transmission line microwave resonator and a movable membrane, simulating the mechanical motion. The membrane is also capacitively coupled to a second transmission line resonator (see Fig. \ref{fig1}). In the strong dispersive limit, the coupling of the superconducting qubit with the resonator and the nanomechanical oscillator gives rise to an effective nonlinear coupling between the resonator and the nanomechanical oscillator. In effect, there are two types of nonlinearities in our system: the nonlinear interaction between the first resonator and the nanomechanical oscillator mediated by the superconducting qubit and the nonlinear interaction induced by the optomechanical coupling between the nanomechanical oscillator and the second microwave resonator. We find that  presence of the superconducting qubit-induced nonlinearity increases the pump power required to observe the bistable behaviour of the mean photon number in the second resonator. We show that the combined effect of these nonlinearities leads to strong squeezing of the transmitted field in the presence of thermal fluctuations. The squeezing is controllable by changing the microwave drive pump power. Using logarithmic negatively as entanglement measure, we also show that the mechanical motion is entangled with the second resonator mode in the steady state. The generated entanglement is shown to be robust against thermal decoherence.

\section{Model and Hamiltonian}
The electromechanical system considered here is schematically depicted in Fig. \ref{fig1}. A superconducting transmission line resonator ($\text{TLR}_{1}$) is placed close to the Cooper-pair box, which is coupled to a large superconducting reservoir via two identical Josephson junctions of capacitance $C_{J}$ and Josephson energy $E_{J}$. This effectively forms a superconducting quantum interface device (SQUID) and is also a basic configuration for superconducting \textit{charge qubit} \cite{Wen08}. The state of the qubit can be controlled by the gate voltage $V_{\textit{g}}$ through a gate capacitance $C_{\textit{g}}$. The qubit is further coupled to a nanomechanical oscillator via capacitance $C_{\rm q}$ that depends on the position x of the membrane (the green line in Fig. \ref{fig1}) from the equilibrium position. Since the amplitude is close to the zero point fluctuation $\text{x}_{\rm zpf}$, the first order correction to the displacement is enough to describe the capacitance. We introduce a dimensionless position operator as $x=\text{x}/\text{x}_{\rm zpf}$, which can be expressed in terms of the annihilation and creation operators as $x=b+b^{\dag}$. Thus, the Hamiltonian of the nanomechanical oscillator of frequency $\hbar\omega_{\rm m}$ is given by $\hbar\omega_{\rm m}(b^{\dag}b+1/2)$ (in our analysis we drop the constant term $\hbar\omega_{\rm m}/2$). If the distance between the membrane and the other arm of the capacitor is $d$ at $\text{x}=0$, then the corresponding capacitance is $C^{(0)}_{q}=\epsilon_{\rm m}S/d$, where $S$ is the surface area of the electrode and $\epsilon_{\rm m}$ is the permittivity of free space. At the displacement $d-\text{x}$ the capacitance reads $C_{q}(\text{x})=C^{(0)}_{q}/(1-\text{x}/d)\simeq C^{(0)}_{q}+C^{(1)}x$, where $C^{(1)}_{q}=x_{\rm zpf}C^{(0)}_{q}/d$. To create a tunable coupling between the microwave resonator and the circuit elements, a gate voltage $V_{q}$ is applied.

The Hamiltonian that describes the interaction of the qubit with the resonator $\text{TLR}_{1}$ and the nanomechanical oscillator, in the rotating wave approximation, is given by \cite{Wen08} (we take $\hbar =1$)
\begin{equation}\label{H}
H_{1}=-\frac{1}{2}\omega_{q} \sigma _{z}+\textit{g}_{c}(c^{\dag}\sigma _{-}+c\sigma
_{+})+\textit{g}_{b}(b^{\dagger 2}\sigma _{-}+b^{2}\sigma _{+}),
\end{equation}
where $\omega_{\rm q}$ is the transition frequency of the qubit, $\textit{g}_{b}$
and $\textit{g}_{c}$ are the microwave resonator-qubit and nanomechanical oscillator-qubit couplings, respectively. The qubit
operators are defined by $\sigma _{z}=\left\vert e\right\rangle \left\langle
e\right\vert -\left\vert g\right\rangle \left\langle g\right\vert ,\sigma
_{+}=(\sigma _{-})^{\dagger }=\left\vert e\right\rangle \left\langle
g\right\vert $ with $\left\vert g\right\rangle $ and $\left\vert
e\right\rangle $ representing the ground and the excited states of the qubit; $b$ and $c$ are the annihilation operators of the mechanical mode and the first resonator $\text{TLR}_{1}$ mode.

Furthermore, the nanomechanical oscillator is coupled to the second transmission line resonator ($\text{TLR}_{2}$), which is externally driven by a microwave field of frequency $\omega_{\rm d}$. This coupling is described by the Hamiltonian
\begin{equation}\label{H2}
H_{2}=\textit{g}_{a}a^{\dag}a(b^{\dag}+b)+i\varepsilon (a^{\dag}e^{-i\omega _{\rm d}t}-ae^{i\omega_{\rm d}t}),
\end{equation}
where $a$ the annihilation operator for the resonator $\text{TLR}_{2}$ mode; $\textit{g}_{a}$
is the resonator-mechanical mode coupling constant, $\varepsilon=\sqrt{2\kappa_{a}P/\hbar \omega_{a}}$ is the amplitude of the microwave drive of $\text{TLR}_{2}$ with $P$ being the corresponding power, $\kappa_{a}$ the resonator damping rate, and $\omega_{a}$ the resonator frequency. The free energies of the mechanical oscillator and the two resonators read
\begin{equation}\label{H0}
H_{0}=\omega_{\rm m}b^{\dag} b+\omega _{a}a^{\dag}a+\omega_{\rm c}c^{\dag}c,
\end{equation}
where $\omega_{\rm m}$ is the mechanical oscillator frequency and $\omega_{\rm c}$ is the frequency of $\text{TLR}_{1}$.

Next, we apply the unitary transformation that effectively eliminates the degrees of freedom of the qubit [in fact the transformation diagonalizes the interaction part of the Hamiltonian \eqref{H}]. This can be achieved by applying a unitary transformation defined by
\begin{equation*}
H=U(H_{0}+H_{1}+H_{2})U^{\dag},
\end{equation*}
where
\begin{equation*}
U=\exp \left[ \frac{g_{c}}{\Delta _{qc}}(c\sigma _{+}-c^{\dag}\sigma _{-})+\frac{%
g_{b}}{\Delta _{qm}}(b^{2}\sigma _{+}-b^{\dagger 2}\sigma _{-})\right],
\end{equation*}
in which $\Delta _{qc}=\omega_{\rm q} -\omega _{c}$ and $\Delta _{qm}=\omega_{\rm q} -2\omega _{\rm m}$. In the dispersive limit,  $\Delta_{qm},\Delta_{qc}\gg\sqrt{g_{b}^2+g_{c}^2}$, the transformation yields an approximate Hamiltonian
\begin{align}\label{HH1}
H &\approx\omega _{a}a^{\dag}a+\omega _{\rm m}b^{\dag}b+\omega _{c}c^{\dag}c +\alpha (b^{\dag 2}c+c^{\dag}b^{2})\sigma_{z}\notag\\
&+\textit{g}_{a}a^{\dag}a(b^{\dag}+b)+i\varepsilon (a^{\dag}e^{-i\omega_{\rm d}t}-ae^{i\omega _{\rm d}t}),
\end{align}
where $\alpha=\textit{g}_{b}\textit{g}_{c}\left( \Delta _{qc}+\Delta _{qm}\right)/2\Delta_{qm}\Delta_{qc}$ is an effective nonlinear coupling between the nanomechanical oscillator and the resonator $\text{TLR}_{1}$. If the qubit is adiabatically kept in the ground state, the effective Hamiltonian reduces to
\begin{align}\label{HH}
H &\approx\omega _{a}a^{\dag}a+\omega _{\rm m}b^{\dag}b+\omega _{c}c^{\dag}c -\alpha (b^{\dag 2}c+c^{\dag}b^{2})\notag\\
&+\textit{g}_{a}a^{\dag}a(b^{\dag}+b)+i\varepsilon (a^{\dag}e^{-i\omega_{\rm d}t}-ae^{i\omega _{\rm d}t}).
\end{align}
Note that if there is strong thermal excitation which promotes the qubit to the excited state, then as follows from \eqref{HH} the sign of the coupling strength obviously change from $-\alpha$ to $\alpha$. The effective nonlinear coupling between the resonator $\text{TLR}_{1}$ and the mechanical mode does not have the same form as the usual optomechanical coupling (e.g., the coupling between $\text{TLR}_{2}$ and the mechanical mode). This is because the former is mediated by a qubit, while the latter is a direct intensity-dependent coupling. 

\subsection{Quantum Langevin equations}
The dynamics of our system can be described by the quantum Langevin equations that take into account the loss of microwave photons from each resonator and the damping of the mechanical motion due to the membrane's thermal bath. In a frame rotating with the microwave drive frequency $\omega_{\rm d}$, the nonlinear quantum Langevin equations read
\begin{equation}\label{a}
\dot {a}=-\left( i\Delta _{a}+\frac{\kappa_{a}}{2}\right)
 a-i\textit{g}_{a} a(b^{\dag}+b)+\varepsilon+\sqrt{\kappa_{a}} a_{\text{in}},
\end{equation}
\begin{equation}\label{b}
\dot b=-(i\omega_{\rm m}+\frac{\gamma _{b}}{2})b-i\text{g}_{a}a^{\dag}a-2i\alpha
cb^{\dag}+\sqrt{\gamma_{m}}b_{\text{in}},
\end{equation}
\begin{equation}\label{c}
\dot c=-\left( i\omega_{c}+\frac{\kappa_{c}}{2}\right) c+i\alpha
b^{2}+\sqrt{\kappa_{c}}c_{\text{in}},
\end{equation}
where $\Delta_{a}=\omega_{a}-\omega_{\rm d}$, and $\kappa_{c}$ and $\gamma_{m}$ are, respectively, the damping rates for the first resonator $\text{TLR}_{1}$ and  mechanical oscillator. We assume that the resonators thermal baths and that of the mechanical bath are Markovian and hence the noise operators $a_{\text{in}}, b_{\text{in}}$, and $c_{\text{in}}$ satisfy the following correlation functions:
\begin{equation}\label{bc1}
  \langle A_{\text{in}}^{\dag}(\omega)A_{\text{in}}(\omega')\rangle=2\pi n_{A}\delta(\omega+\omega'),
\end{equation}
\begin{equation}\label{bc2}
  \langle A_{\text{in}}(\omega)A_{\text{in}}^{\dag}(\omega')\rangle=2\pi (n_{A}+1)\delta(\omega+\omega'),
\end{equation}
with $n_{A}^{-1}=\exp(\hbar\omega_{A}/k_{B}T_{A})-1$, where $k_{B}$ is the Boltzmann constant and $A=a,b,c$, and the noise operators have zero-mean values, $\langle a_{\rm in}\rangle=\langle b_{\rm in}\rangle=\langle c_{\rm in}\rangle=0$.

\subsection{Optical bistability in resonator photon number}
It is well-known that for strong enough pump power and in the red-detuned ($\omega_{d}-\omega_{a}<0$) regime, an optomechanical coupling gives rise to optical bistability. Here we investigate the effect of the nonlinearity induced by the superconducting qubit on the bistable behaviour. Solving the expectation values of Eqs. \eqref{a}-\eqref{c} in the steady state we obtain
\begin{equation}\label{aave}
  \langle a\rangle=\frac{\varepsilon}{i\Delta_{\rm f}+\kappa_{a}/2},
\end{equation}
\begin{equation}\label{bave}
  \langle b\rangle=\frac{-ig_{a}|\langle a\rangle|^2}{i\omega_{\rm m}+\gamma_{\rm m}/2}-i\frac{2\alpha\langle c\rangle\langle b^{\dag}\rangle}{i\omega_{\rm m}+\gamma_{\rm m}/2},
\end{equation}
\begin{equation}\label{cave}
  \langle c\rangle=i\frac{\alpha\langle b\rangle^2}{i\omega_{c}+\kappa_{c}/2},
\end{equation}
where $\Delta_{\rm f}=\Delta_{a}+g_{a}(\langle b\rangle+\langle b^{\dag}\rangle)$ is an effective detuning for second resonator. Combining these equations, we obtain the coupled equations for the mean photon number $I_{a}=|\langle a\rangle|^2$ in the second resonator and the mean phonon number $I_{b}=|\langle b\rangle|^2$ as
\begin{figure}[t]
\includegraphics[width=4.4cm,height=4.4cm]
{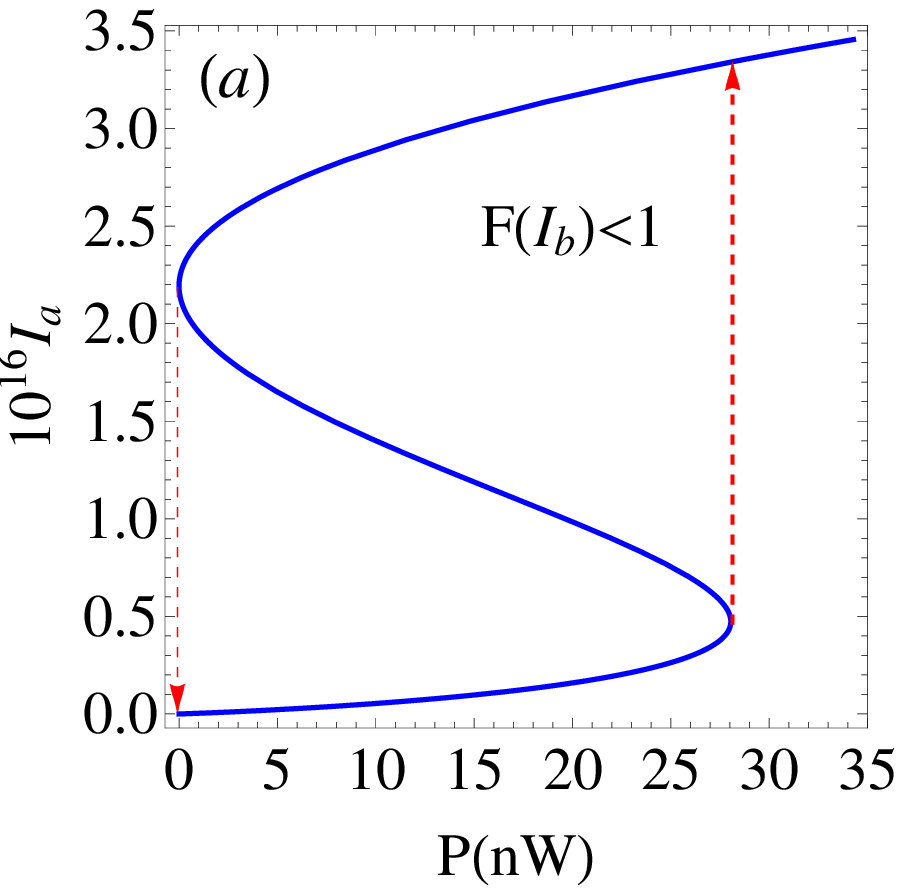}\includegraphics[width=4.3cm,height=4.3cm]{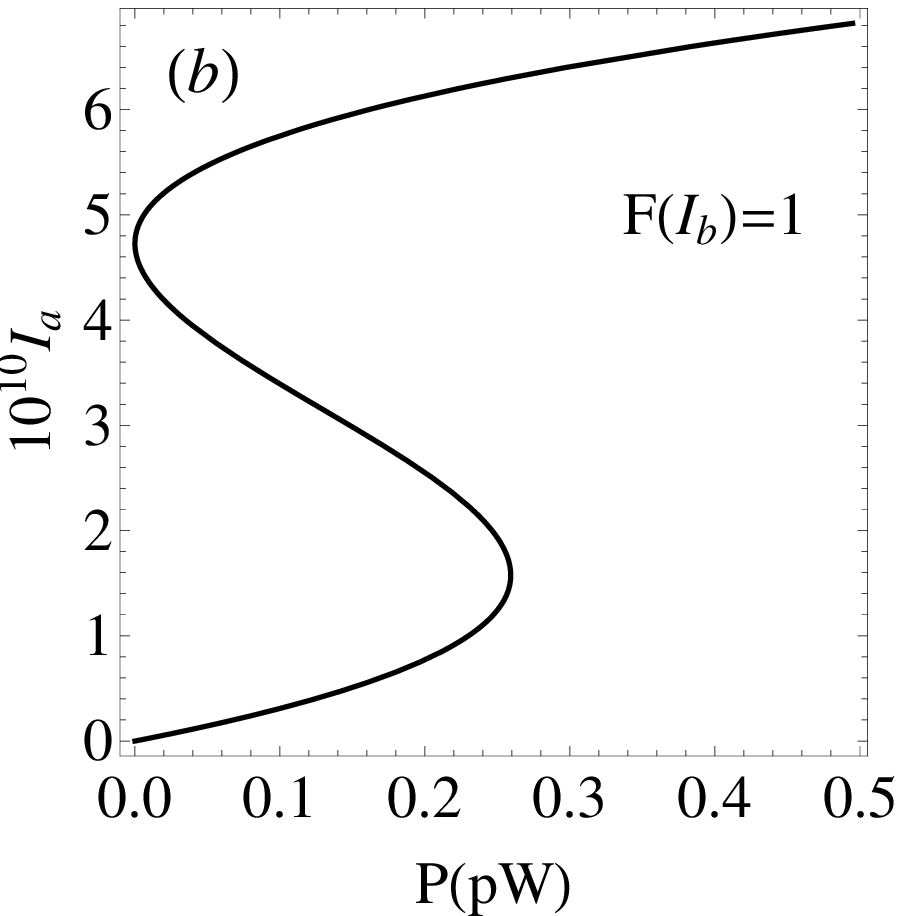}
\caption{Bistability behaviour for mean photon number in the second resonator $I_{a}$ (a) in the presence of the nonlinear coupling $\alpha\neq 0$ [$F(I_{b})<1$] (b) in the absence of the nonlinear coupling $\alpha=0$[$F(I_{b})=1$]. The parameters used are: frequencies $\omega_{\rm m}/2\pi=10~\text{MHz}$, $\omega_{a}/2\pi=7.5~\text{GHz}$, $\omega_{c}/2\pi=2.5~\text{GHz}$, $\omega_{q}/2\pi=3~\text{GHz}$, $\omega_{d}/2\pi=7~\text{GHz}$, couplings  $g_{a}/\pi=460\text{Hz}$,$g_{b}/2\pi=2\text{MHz}$, $g_{c}/2\pi=30~\text{MHz}$, and damping rates $\kappa_{a}/2\pi=10^{5}~\text{Hz}$, $\gamma_{\rm m}/2\pi=50~\text{Hz}$, and $\kappa_{c}/2\pi=10^{5}~\text{Hz}$. }\label{fig2}
\end{figure}
\begin{equation}\label{bis}
  I_{a}\left[\left(\Delta_{a}-F(I_{b})\frac{2\textit{g}_{a}^2\omega_{\rm m}I_{a}}{\omega_{\rm m}^2+(\gamma_{\rm m}/2)^2}\right)^2+\left(\frac{\kappa_{a}}{2}\right)^2\right]=|\varepsilon|^2,
\end{equation}
\begin{align}\label{f2}
&I_{a}^2=\notag\\
&\frac{I_{b}[(1+I_{b}\beta_{1})^2+I_{b}^2\beta_{2}^2]^2[\omega_{\rm m}^2+(\gamma_{\rm m}/2)^2]^2/g_{a}^2}{[\omega_{\rm m}(1+I_{b}\beta_{1})+\frac{\gamma_{\rm m}}{2}I_{b}\beta_{2}]^2+[\frac{\gamma_{\rm m}}{2}(1+I_{b}\beta_{1})-\omega_{\rm m}I_{b}\beta_{2}]^2}
\end{align}
where
\begin{equation}\label{chi}
 F(I_{b})=\frac{1+I_{b}\beta_{1}+\frac{\gamma_{\rm m}}{2\omega_{\rm m}}I_{b}\beta_{2}}{(1+I_{b}\beta_{1})^2+I_{b}^2\beta_{2}^2},
\end{equation}
\begin{equation}\label{chi1}
 \beta_{1}=\frac{2\alpha^2(\omega_{\rm m}\omega_{c}-\gamma_{\rm m}\kappa_{c}/4)}{[\omega_{\rm m}^2+(\gamma_{\rm m}/2)^2][\omega_{c}^2+(\kappa_{c}/2)^2]},
\end{equation}
\begin{equation}\label{chi2}
 \beta_{2}=\frac{\alpha^2(\omega_{\rm m}\kappa_{c}+\omega_{c}\gamma_{\rm m})}{[\omega_{\rm m}^2+(\gamma_{\rm m}/2)^2][\omega_{c}^2+(\kappa_{c}/2)^2]}.
\end{equation}
\begin{figure}[t]
\includegraphics[width=7cm]{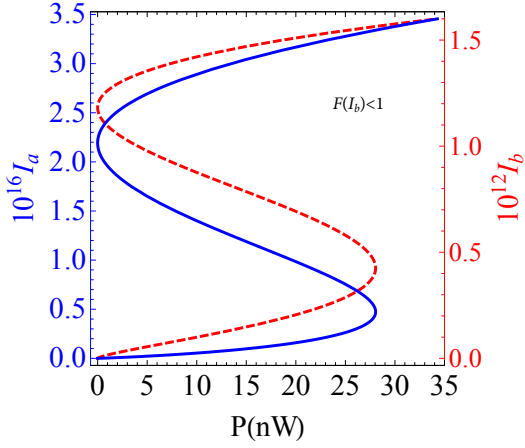}
\caption{Bistability behaviour for mean photon number in the second resonator $I_{a}$ (blue solid curve) and mean photon number $I_{b}$ (red dashed curve) as a function of the pump power in the presence of the superconducing qubit ($\alpha\neq 0, F(I_{b}<1)$). All parameters are the same as in Fig. \ref{fig2}.}\label{fig3}
\end{figure}
We immediately see from Eq. \eqref{bis} that in the absence of the superconducting circuit, which amounts to setting $\alpha=0$ in \eqref{chi1} and \eqref{chi2}, the factor $F$ that appears in \eqref{bis} becomes, $F(I_{b})=1$. The resulting equation reproduces the cubic equation for the mean photon number $I_{a}$ as in the standard optomechanical coupling \cite{Set12}, which is known to exhibit bistable behaviour. In general, for electromechanical system considered here, $F(I_{b})<1$ (for typical experimental parameters \cite{Teu11}), thus yielding the same form of cubic equation for $I_{a}$. In Fig. \ref{fig2}, we plot the mean photon number $I_{a}$ as function of the pump power in the presence and absence of the superconducting qubit. Figure \ref{fig2}a shows, in the presence of the qubit ($\alpha\neq 0$), the bistability behaviour only appears when the microwave resonator is pumped at nW range. For example, for the parameters used in Fig. \eqref{fig2}a, the lower tuning point is obtained at $P\approx 28\text{nW}$. The hysteresis then follows the arrow and jumps to the upper branch. Then scanning the pump power towards zero, one obtains the other turning point at very low pump power $P=0.02\text{pW}$. On the other hand, in the absence of the superconducting qubit (see Fig. \ref{fig2}b), the pump power required to achieve the bistable behaviour reduces to the pW range, with the lower turning point appearing at $P=0.26\text{pW}$. Therefore, the bistable behaviour in the mean photon number in the second resonator can be observed at relatively high pump power when the nanomechanical oscillator coupled to the superconducting qubit.Therefore, when the nanomechanical oscillator is coupled to the superconducting qubit, a relatively high
power is needed to observe a bistable behavior.

Furthermore, according to Eq. \eqref{f2}, since $\alpha/\omega_{\rm m}\ll1(\beta_{i}\approx 0)$, the mean photon number $I_{a}$ is related to the phonon number via $I_{a}^2=I_{b}[\omega_{\rm m}^2+(\gamma_{\rm m}/2)^2]/g_{a}^2$, indicating that the phonon number also exhibits bistability. Figure \ref{fig3} compares the bistable behavior for both $I_{a}$ and $I_{b}$. As can be seen from this figure, the bistability occurs at the same power range; however, their corresponding photon and phonon numbers are different by four orders of magnitude. Note that, as expected, all the bistable behaviours are observed in the red detuned regime, $\Delta_{a}=\omega_{a}-\omega_{d}>0$.  From application viewpoint, the bistable behaviour can used as a fast optical switching.

\subsection{Fluctuations about the classical mean value}
The quantum Langevin equations [Eqs. \eqref{a}-\eqref{c}] can be solved analytically by adopting a linearization scheme \cite{Set10,Set11} in which the operators are expressed as the sum of their mean values plus fluctuations, that is, $a=\langle a\rangle +\delta a$, $b=\langle b\rangle+\delta b$, and $c=\langle c\rangle+\delta c$. The equations for fluctuation operators then read
\begin{align}\label{da}
 \delta \dot a = -\left( i\Delta_{\rm f}+\frac{\kappa_{a}}{2}\right)\delta a-i\textit{g}_{a}\langle a\rangle(\delta b+\delta b^{\dag})+\sqrt{\kappa_{a}}a_{\text{in}},
 \end{align}
 \begin{align}\label{db}
\delta \dot b =& -\left(i\omega_{\rm m}+\frac{\gamma_{\rm m}}{2}\right)\delta b-i\textit{g}_{a}(\langle a^{\dag}\rangle \delta a+\langle a\rangle \delta a^{\dag})\notag\\
 &-2i\alpha[\langle c\rangle\delta b^{\dag}+\langle b^{\dag}\rangle\delta c]+ \sqrt{\gamma_{\rm m}}b_{\text{in}},
\end{align}
\begin{align}\label{dc}
 \delta \dot c=-\left( i\omega_{c}+\frac{\kappa_{c}}{2}\right) \delta c+2i\alpha
\langle b\rangle \delta b+\sqrt{\kappa_{c}}c_{\text{in}}.
\end{align}
The solutions to these equations can easily be obtained in frequency domain. To this end, writing the Fourier transform of Eqs. \eqref{da}-\eqref{dc} and their complex conjugates, we get
\begin{equation}
\mathcal{A}\mathcal{U}=\mathcal{N},
\end{equation}
where
\begin{equation}\label{MM}
 \mathcal{A}=\left(
      \begin{array}{cccccc}
        \eta_{+} & 0 & G & G & 0 & 0 \\
        0 & \eta_{-} & G^{*} & G^{*} & 0 & 0 \\
        -G^{*} &G & v_{+} &\mathcal{C} & \mathcal{B}^{*} & 0 \\
        G^{*} & -G & \mathcal{C}^{*}& v_{-} & 0 & \mathcal{B} \\
        0 & 0 & \mathcal{B} & 0 & u_{+} &0 \\
        0 & 0 & 0 & \mathcal{B}^{*}  & 0 & u_{-} \\
      \end{array}
    \right),
\end{equation}
$\mathcal{U}=(\delta a,\delta a^{\dag},\delta b,\delta b^{\dag},\delta c,\delta c^{\dag})^{T}$ and $\mathcal{N}=(\sqrt{\kappa_{a}}a_{\rm in}, \sqrt{\kappa_{a}}a_{\rm in}^{\dag}, \sqrt{\gamma_{\rm m}}b_{\rm in},\sqrt{\gamma_{\rm m}}b_{\rm in}^{\dag},\sqrt{\kappa_{c}}c_{\rm in},\sqrt{\kappa_{c}}c_{\rm in} ^{\dag})^{T}$ with $ \eta_{\pm}=\kappa_{a}/2+ i(\omega \pm \Delta_{\rm f})$, $v_{\pm}=\gamma_{\rm m}/2+i(\omega\pm\omega_{\rm m})$, and $  u_{\pm}=\kappa_{c}/2+i(\omega\pm\omega_{c})$,
 $ G=i\textit{g}_{a}\langle a\rangle,\mathcal{B}=-2i\alpha\langle b\rangle,\mathcal{C}=2i\alpha\langle c\rangle.$

The solution for the fluctuation operator $\delta a$ of the second resonator field has the form
\begin{equation}\label{sda}
  \delta a(\omega)=\xi_{1}a_{\text{in}}+\xi_{2}a_{\text{in}}^{\dag}+\xi_{3}b_{\text{in}}+\xi_{4}b_{\text{in}}^{\dag}+\xi_{5}c_{\text{in}}+\xi_{6}c_{\text{in}}^{\dag}.
\end{equation}
The explicit expression for the coefficients $\xi_{i}$ are given in the Appendix. Similarly, the expressions for $\delta b(\omega)$ and $\delta c(\omega)$ can be obtained from \eqref{MM}. In the following, we use \eqref{sda} to analyze the squeezing of the transmitted microwave field from the second resonator.

\section{Squeezing spectrum}
It was shown that the optomechanical coupling can lead to squeezing of the nanomechanical motion, which can be inferred by measuring the squeezing of the transmitted microwave field \cite{Fab94,Woo08,Set11}. Here we investigate the squeezing properties of the transmitted microwave field in the presence of the nonlinearity induced by superconducting qubit [represented by the effective coupling $\alpha$ in Eq. \eqref{HH}] as well as the nonlinearity due to the optomechanical coupling [represented by coupling $\textit{g}_{a}$ in Eq. \eqref{HH}]. The stationary squeezing spectrum of the transmitted field is given by
\begin{align}\label{Sp}
  S(\omega)&=\int_{-\infty}^{\infty}d\tau \langle \delta X_{\phi}^{\text{out}}(t+\tau)\delta X_{\phi}^{\text{out}}(t)\rangle_{\text{ss}}e^{i\omega\tau}\notag\\
  &=\langle \delta X_{\phi}^{\text{out}}(\omega)\delta X_{\phi}^{\text{out}}(\omega)\rangle
\end{align}
where $\delta X_{\phi}^{\text{out}}=e^{i\phi}\delta a_{\text{out}}+e^{-i\phi}\delta a^{\dag}_{\text{out}}$ with $a_{\text{out}}=\sqrt{\kappa_{a}}\delta a-a_{\text{in}}$ being the input-output relation \cite{Mil94} and $\phi$ the measurement phase angle determined by the local oscillator. The squeezing spectrum can be put in the form
\begin{equation}\label{SP1}
  S(\omega)=1+C_{a^{\dag}a}^{\text{out}}+e^{-2i\phi}C_{aa}^{\text{out}}+e^{2i\phi}C_{a^{\dag}a^{\dag}}^{\text{out}},
\end{equation}
where $\langle \delta a_{\text{out}}(\omega)\delta a_{\text{out}}(\omega')\rangle=2\pi C_{aa}^{\text{out}}(\omega)\delta(\omega+\omega')$ and $\langle \delta a_{\text{out}}(\omega)^{\dag}\delta a_{\text{out}}(\omega')\rangle=2\pi C_{a^{\dag}a}^{\text{out}}(\omega)\delta(\omega+\omega')$. The degree of squeezing depends on the direction of the measurement of the quadratures, thus can be optimized over the phase angle $\phi$. Using the angle which optimizes the squeezing \cite{Set111}, we obtain
\begin{equation}\label{sq}
  S_{\text{opt}}^{(\pm)}(\omega)=1+2C_{a^{\dag}a}^{\text{out}}(\omega)\pm 2|C_{aa}^{\text{out}}(\omega)|.
\end{equation}
$S_{\text{opt}}^{(-)}$ corresponds to the spectrum of the squeezed quadrature, while $S_{\text{opt}}^{(+)}$ represents the spectrum of the unsqueezed quadrature. Using the solution \eqref{sda} and the correlation properties of the noise forces \eqref{bc1} and \eqref{bc2}, the spectrum of the squeezed quadrature takes the form
\begin{align}\label{Sopt}
S_{\text{opt}}^{(-)}(\omega)=1+2C_{a^{\dag}a}^{\text{out}}(\omega)- 2|C_{aa}^{\text{out}}(\omega)|,
\end{align}
where
\begin{align}\label{cada}
C_{a^{\dag}a}^{\text{out}}(\omega)=&\kappa_{a}\big[n_{a}\xi_{1}(\omega)\xi_{1}^{*}(-\omega)+(n_{a}+1)\xi_{2}(\omega)\xi_{2}^{*}(-\omega)\notag\\
&+n_{b}\xi_{3}(\omega)\xi_{3}^{*}(-\omega)+(n_{b}+1)\xi_{4}(\omega)\xi_{4}^{*}(-\omega)\notag\\
&+n_{c}\xi_{5}(\omega)\xi_{5}^{*}(-\omega)+(n_{c}+1)\xi_{6}(\omega)\xi_{6}^{*}(-\omega)\big]\notag\\
&-2\sqrt{\kappa_{a}}n_{a}[\xi_{1}(\omega)+\xi_{1}^{*}(-\omega)]+n_{a},
\end{align}
\begin{align}\label{cada}
C_{aa}^{\text{out}}(\omega)=&\kappa_{a}\big[n_{a}\xi_{1}(\omega)\xi_{2}^{*}(-\omega)+(n_{a}+1)\xi_{1}^{*}(-\omega)\xi_{2}(\omega)\notag\\
&+n_{b}\xi_{3}(\omega)\xi_{4}^{*}(-\omega)+(n_{b}+1)\xi_{3}^{*}(-\omega)\xi_{4}(\omega)\notag\\
&+n_{c}\xi_{5}(\omega)\xi_{6}^{*}(-\omega)+(n_{c}+1)\xi_{5}^{*}(-\omega)\xi_{6}(\omega)\big]\notag\\
&-\sqrt{\kappa_{a}}[n_{a}\xi_{2}^{*}(-\omega)+(n_{a}+1)\xi_{2}(\omega)].
\end{align}
Based on the definition of the quadrature $\delta X_{\varphi}^{\text{out}}$, the microwave field is squeezed when the value of the squeezing spectrum is below the standard quantum limit, $S_{\text{opt}}^{(-)}(\omega)=1$.

\begin{figure}[t]
\includegraphics[width=8cm]{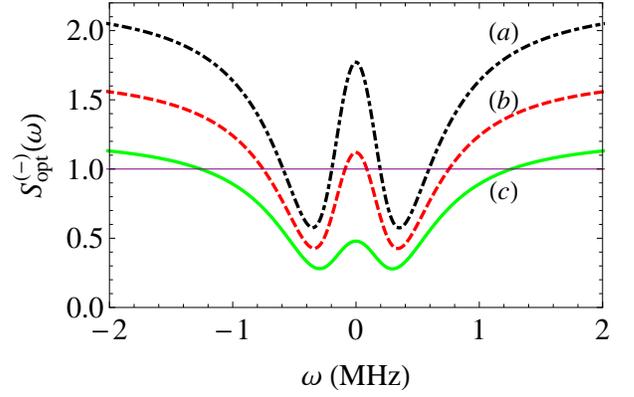}
\caption{Plots of the squeezing spectrum of the transmitted microwave field [Eq. \eqref{Sopt}] for drive pump power $P=8~\mu\text{W}$, for drive frequency $\omega_{d}/2\pi=7.4999~\text{GHz}$, for membrane's bath temperature $T_{b}=50~\text{mK}$, for bath temperature of the first resonator, $T_{c}=2~\text{K}$, and for various bath temperatures of the second resonator: (a) $T_{a}=150~\text{mK}$ (solid green curve), (b) $T_{b}=250~\text{mK}$ (dashed red curve), and (c) $T_{b}=350~\text{mK}$ (dot-dashed black curve). The horizontal solid line represents the standard quantum limit [$S_{\text{opt}}^{(-)}(\omega)=1$], below which indicates squeezing. All other parameters are the same as in Fig. \ref{fig2}.}\label{fig4}
\end{figure}

In Fig. \ref{fig4}, we plot the squeezing spectrum of the microwave field as a function of the temperature $T_{a}$ of the second resonator thermal bath. As can be seen from this figure, the microwave field exhibits squeezing with the degree of squeezing strongly relying on the thermal bath temperature, $T_{a}$. Obviously, the amount of squeezing degrades as the thermal temperature increases and it ultimately disappears when the bath temperature reaches at $T_{a}\approx600~\text{mK}$ for the parameters used in Fig. \ref{fig4}. We also found that the degree of squeezing is less sensitive to the first resonator thermal bath temperature $T_{c}$. This is because the second resonator is not directly coupled to the first resonator thermal bath, though it is indirectly coupled via the nanomechanical oscillator through a low-loss capacitor. The other interesting aspect is that the spectrum shows double dips for strong enough pump power indicating that the optomechanical interaction reached the strong coupling regime, a requirement to observe quantum mechanical effects. It is worth mentioning that to make sure that the squeezing is determined in the stable regime, the microwave drive frequency $\omega_{d}$ is deliberately chosen close to resonance frequency of the second resonator $\omega_{a}$.
\begin{figure}[t]
\includegraphics[width=7.5cm]{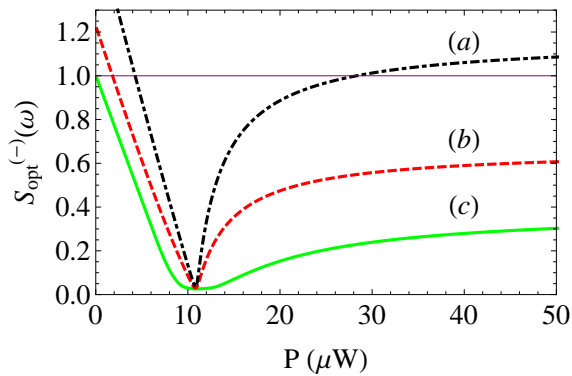}
\caption{Plots of the squeezing spectrum vs the microwave drive pump power P ($\mu$W) for the bath temperature of the first resonator $T_{c}=2~\text{K}$, the membrane's bath temperature, $T_{b}=10~\text{mK}$, and for different values of the bath temperature $T_{a}$ of the second resonator: (a) $T_{a}=250~ \text{mK}$ (dot-dashed black curve), (b) $T_{a}=150~\text{mK}$ (dashed red curve), and (c) $T_{a}=50~ \text{mK}$ (solid green curve). All other parameters are the same as in Fig. \ref{fig2}.}\label{fig5}
\end{figure}
\begin{figure}[t]
\includegraphics[width=7.5cm]{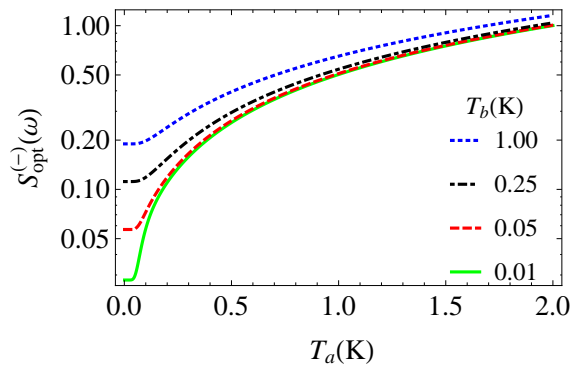}
\caption{Plots of the squeezing spectrum (in logarithmic scale) vs the bath temperature of the second resonator $T_{a}$ for a pump power $P=10~\mu\text{W}$, for the bath temperature $T_{c}=2~\text{K}$ of the first resonator, and for different values of the membrane's bath temperature, $T_{b}=1~\text{K}$ (dotted blue curve), $T_{b}=0.25~\text{K}$ (dot-dashed black curve), $T_{b}=0.05~\text{K}$ (dashed red curve), $T_{b}=0.01~\text{K}$ (solid green curve). All other parameters are the same as in Fig. \ref{fig2}.}\label{fig66}
\end{figure}

The other important external parameter that can be used to control the degree of the squeezing is the strength of the microwave drive. The dependence of the squeezing on the drive pump power is illustrated in Fig. \ref{fig5}. When the microwave drive frequency is close to the resonator frequency, that is, when $\Delta_{a}/2\pi=0.1\text{MHz}$, the squeezing gradually develops as the pump power is increased to the range of few $\mu \text{W}$. Further increase in the pump power strength leads to an optimum squeezing that can possibly be achieved for a given value of temperature of the thermal baths. For example, for $T_{a}=10\text{mK}, T_{b}=10\text{mK}$, and $T_{c}=2\text{K}$, the maximum squeezing is $\approx 97\%$ below the standard quantum limit at a pump power $P\approx 10\mu\text{W}$. However, when the pump power is increased beyond $P\approx 10\text{mW}$, the degree of squeezing sharply decrease and becomes strongly dependent on $T_{a}$. The other interesting aspect is that although the bath temperature $T_{a}$ is increased to $250 \text{mK}$, there exists an optimum power for which the squeezing is still the maximum achievable. Even though the overall squeezing is due to both nonlinearities induced by the effective coupling between the first resonator and the nanomechanical oscillator and the optomechanical coupling, the enhancement of the squeezing with pump power is mainly due to the optomechanical coupling. This is because the pump power directly affects the intensity in the second resonator ($\text{TLR}_{2}$), which in turn increases the strength of the optomechanical coupling.

Fixing the power ($P=10 \text{mW}$) at which the squeezing is maximum, it is important to understand the interplay between the bath temperatures $T_{a}$ and $T_{b}$ in determining the degree of squeezing of the microwave field. Figure \ref{fig66} shows that the squeezing persists up to $T_{a}\approx \text{2K}$. While the degree of squeezing is weakly dependent on the thermal bath temperature $T_{b}$ of the nanomechanical oscillator when $T_{a}>0.1\text{K}$, the squeezing decreases with increasing $T_{b}$ for $T_{a}<0.1\text{K}$. Therefore, a strong and robust squeezing can be achieved by tuning the pump power close to $P=10 \mu\text{W}$ while keeping the bath temperatures $T_{a},T_{b}$ within $\lesssim 1$K range.

 \section{Optomechanical entanglement}
It has been shown that the optomechanical coupling gives rise to entanglement between the resonator field and mechanical motion \cite{Zha03,Pin05,Vit08}. Here we analyze the robustness of the optomechanical entanglement against thermal decoherence in the presence of the two different nonlinearities discussed earlier. We also analyze how the degree entanglement depends on the drive pump power and the detuning $\Delta_{a}$. In order to investigate the optomechanical entanglement, it is more convenient to use the quadrature operators defined by
\begin{align}
 & X_{a} = \frac{1}{\sqrt{2}}(\delta a+\delta a^{\dag}), Y_{a} = \frac{1}{\sqrt{2}i}(\delta a-\delta a^{\dag}),  \\
 & X_{b} = \frac{1}{\sqrt{2}}(\delta b+\delta b^{\dag}), Y_{b} = \frac{1}{\sqrt{2}i}(\delta b-\delta b^{\dag}), \\
 & X_{c} = \frac{1}{\sqrt{2}}(\delta c+\delta c^{\dag}), Y_{c} = \frac{1}{\sqrt{2}i}(\delta c-\delta c^{\dag}).
\end{align}
The equations of motion for these quadrature operators can be put in a matrix form
\begin{equation}\label{M}
  \dot u(t) =Mu(t)+f(t),
\end{equation}
where
\begin{widetext}
\begin{equation}\label{AA}
  R=\left(
      \begin{array}{cccccc}
        -\kappa_{a}/2 & \Delta_{f\rm} & -2\text{g}_{a}\eta_{b} & 0 & 0 & 0 \\
        \Delta_{f\rm} & -\kappa_{a}/2 & -2\text{g}_{a}\mu_{a} & 0 & 0 & 0 \\
        0 &0 &-\gamma_{\rm m}/2+2\alpha\mu_{c} & \omega_{\rm m}-2\alpha\eta_{c}&-2\alpha\mu_{b}& 2\alpha\eta_{b} \\
        -2\text{g}_{a}\eta_{a} & -2\text{g}_{a}\mu_{a}& -(\omega_{\rm m}+2\alpha\eta_{c}) & -(\gamma_{\rm m}/2+2\alpha\mu_{c}) & -2\alpha\eta_{b}&-2\alpha\mu_{b} \\
        0 & 0 & -2\alpha\mu_{b} & -2\alpha\eta_{b} & -\kappa_{c}/2 &\omega_{c} \\
        0 & 0 & 2\alpha\eta_{b} & -2\alpha\mu_{b}  & -\omega_{c} & -\kappa_{c}/2 \\
      \end{array}
    \right), u=\left(\begin{array}{c}
     \delta X_{a} \\
      \delta Y_{a} \\
      \delta X_{b} \\
      \delta Y_{b}\\
      \delta X_{c} \\
      \delta Y_{c}
    \end{array}\right)
    , f=\left(\begin{array}{c}
      \sqrt{\kappa_{a}} X_{a}^{\text{in}} \\
      \sqrt{\kappa_{a}} Y_{a}^{\text{in}} \\
      \sqrt{\gamma_{\rm m}} X_{b}^{\text{in}} \\
      \sqrt{\gamma_{\rm m}} Y_{b}^{\text{in}}\\
      \sqrt{\kappa_{c}} X_{c}^{\text{in}} \\
      \sqrt{\kappa_{c}} Y_{c}^{\text{in}}
    \end{array}\right),
    \end{equation}
\end{widetext}
where $\eta_{L}=\frac{1}{2}(\langle L\rangle+\langle L^{\dag}\rangle)$, $\mu_{L}=\frac{1}{2i}(\langle L\rangle-\langle L^{\dag}\rangle)$ and $X_{L}^{\text{in}}=(\delta L_{\text{in}}+\delta L^{\dag}_{\text{in}})/\sqrt{2}$, $Y_{L}^{\text{in}}=i(\delta L^{\dag}_{\text{in}}-\delta L_{\text{in}})/\sqrt{2}$, where $L=a,b,c$.

In this work, we are interested in the steady state optomechanical entanglement. It is then sufficient to focus on the subspace spanned by the second resonator and mechanical mode (the upper left $4\times 4$ matrix in $R$). To study the stationary optomechanical entanglement, one needs to find a stable solution for Eq. \eqref{M}, so that it reaches a unique steady state independent of the initial condition. Since we have assumed the quantum noises $a_{\rm in}, b_{\rm in}$ and $c_{\rm in}$ to be zero-mean Gaussian noises and the corresponding equations for fluctuations $(\delta a, \delta b$, and $\delta c$) are linearized, the quantum steady state for fluctuations is simply a zero-mean Gaussian state, which is fully characterized by $4\times 4$ correlation matrix $V_{ij}=[\langle u_{i}(\infty)u_{j}(\infty)+u_{j}(\infty)u_{i}(\infty)\rangle]/2$. The solution to Eq. \eqref{M}, $u(t)=M(t)u(0)+\int_{0}^{t}dt'M(t')f(t-t')$, where $M(t)=\exp(Rt)$,  is stable and reaches steady state when all of the eigenvalues of $R$ have negative real parts so that $M(\infty)=0$. The stability condition can be derived by applying the Routh-Hurwitz criterion \cite{DeJ87}. For all results presented in this paper, the stability conditions are satisfied. When the system is stable one easily get
\begin{equation}\label{st}
 \mathcal{V}_{ij}=\sum_{lm}\int_{0}^{\infty}dt'\int_{0}^{\infty}dt''M_{il}(t')M_{jm}\Pi_{lm}(t'-t''),
  \end{equation}
where the stationary noise correlation matrix is give by $\Pi_{lm}=[\langle  f_{l}(t)f_{m}(t'')+ f_{m}(t'')f_{l}(t)\rangle]/2$, where $f_{i}$ is the $i$th element of the column vector $f$. Since all noise correlations are assumed to be Markovian (delta-correlated) and all components of $f(t)$ are uncorrelated, the noise correlation matrix takes a simple form $\Pi_{lm}(t'-t'')=D_{lm}\delta(t'-t'')$, where
\begin{align}
D=&\text{Diag}[\kappa_{a}(2n_{a}+1)/2,\kappa_{a}(2n_{b}+1)/2,\gamma_{\rm m}(2n_{b}+1)/2,\notag\\
&\gamma_{\rm m}(2n_{b}+1)/2,\kappa_{c}(2n_{c}+1)/2,\kappa_{c}(2n_{c}+1)/2]
\end{align}
is the diagonal matrix. As a result, Eq. \eqref{st} becomes $V=\int_{0}^{\infty}dt'M(t')DM(t')^{\rm T}$. When the stability conditions are satisfied, i.e., $M(\infty)=0$, one readily obtain an equation for steady state correlation matrix
\begin{equation}\label{cm}
  R\mathcal{V}+\mathcal{V}R^{\rm T}=-D.
\end{equation}
Equation \eqref{cm} is a linear equation (also known as Lyapunov equation) for $\mathcal{V}$ and can be solved in straight-forward manner. However, the solution for our system is rather lengthy and will not be presented here. We instead solve \eqref{cm} numerically to analyze the optomechanical entanglement.

In order to analyze the optomechanical entanglement, we employ the logarithmic negativity $E_{N}$, a quantity which has been proposed as a measure of bipartite entanglement \cite{Vid02}. For continuous variables, $E_{N}$ is defined as
\begin{equation}\label{LN}
  E_{N}=\max [0,-\ln 2\chi],
\end{equation}
where $\chi=2^{-1/2}\left[\sigma-\sqrt{\sigma^2-4\text{det} \mathcal{V}}\right]^{1/2}$ is the lowest simplistic eigenvalue of the partial transpose of the $4\times 4$ correlation matrix $\mathcal{V}$ with $\sigma=\det \mathcal{V}_{A}+\det \mathcal{V}_{B}-2\det \mathcal{V}_{AB}$. Here $\mathcal{V}_{A}$ and $\mathcal{V}_{B}$ represent the second resonator field and mechanical mode, respectively, while $\mathcal{V}_{AB}$ describes the optomechanical correlation.  These matrices are elements of the $2\times2$ block form of the correlation matrix
\begin{equation}\label{vv}
  \mathcal{V}\equiv \left(
                      \begin{array}{cc}
                        \mathcal{V}_{A} & \mathcal{V}_{AB}\\
                       \mathcal{V}_{AB}^{T} & \mathcal{V}_{B} \\
                      \end{array}
                    \right).
\end{equation}
Any two modes are said to be entangled when the logarithmic negativity $E_{N}$ is positive.
\begin{figure}[t]
  \includegraphics[width=7.5cm]{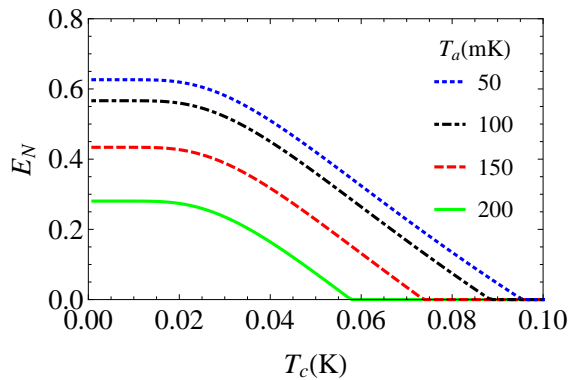}
  \caption{Plots of the logarithmic negativity $E_{N}$ vs the temperature of the first resonator thermal bath, $T_{c}$ for the drive pump power $P=1~\mu\text{W}$, $\Delta_{a}/2\pi=0.1\text{MHz}$ and for different values of the second resonator thermal bath temperature $T_{a}$= $50~\text{mK}$ (dotted blue curve), $100~\text{mK}$ (dotdashed black curve), $150~\text{mK}$ (dashed red curve), and $200~\text{mK}$ (solid green curve). All other parameters the same as in Fig. \ref{fig2}.}\label{fig6}
\end{figure}

\begin{figure}[h]
  \includegraphics[width=7.5cm]{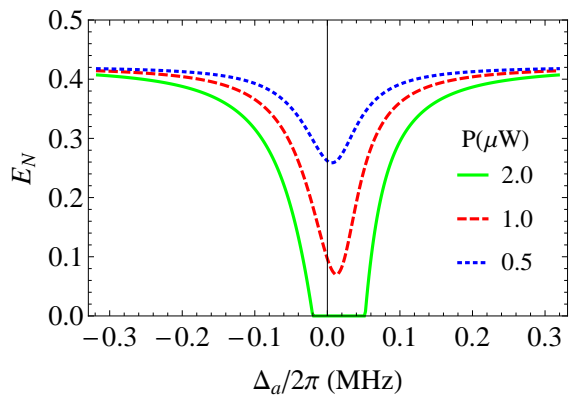}
  \caption{Plots of the logarithmic negativity $E_{N}$ vs the detuning $\Delta_{a}$  for the thermal bath temperature of the first resonator $T_{c}=50~\text{mK}$ and for the thermal bath temperature of the second resonator $T_{a}$= $100~\text{mK}$ and for different values of the microwave drive pump power $P=0.5~\mu\text{W}$ (dotted blue curve), $1.0~\mu\text{W}$ (dashed red curve), and $2.0~ \mu\text{W}$ (solid green curve). All other parameters as the same as in Fig. \ref{fig2}. }\label{fig8}
\end{figure}
In Fig. \ref{fig6}, we plot the logarithmic negativity $E_{N}$ as a function the thermal bath temperature $T_{c}$ of the first resonator while varying the thermal bath temperature $T_{a}$ of the second resonator at a fixed drive pump power, $P=1 \mu\text{W}$. This figure shows that the mechanical mode is entangled with the resonator mode of the second resonator in the steady state. The entanglement strongly relies on the bath temperatures $T_{a}$ and $T_{c}$ of the first and second resonators, respectively. In general, the optomechanical entanglement degrades as the thermal bath temperatures increases. For instance, when the temperature of the second resonator fixed at $50\text{mK}$, the entanglement survives until the bath temperature $T_{c}$ of the first resonator reaches about $ 100\text{mK}$. If the temperature $T_{a}$ is further increased, the critical temperature $T_{c}$ above which the entanglement disappears decreases. Therefore, at constant pump power, the entanglement can be controlled by tuning the bath temperatures of the two resonators.


Another system parameter that can be used as an external knob to control the degree of entanglement is the detuning $\Delta_{a}$. Figure \ref{fig8} illustrates the logarithmic negativity versus the detuning $\Delta_{a}$ for different values of the pump power. Close to resonance ($\Delta_{a}=0$) and for the pump power $P\gtrsim 1.2 \mu\text{W}$, there is no optomechanical entanglement; however, the entanglement between the nanomechanical oscillator and the resonator field arises when the detuning is further increased, and reaches stationary values for $\Delta_{a}/2\pi\simeq \omega_{\rm m}/2\pi=10\text{MHz}$, which is consistent with the results in the literature \cite{Vit08}. The interesting aspect of our result is that the entanglement persists for wide range of detuning $\Delta_{a}$, opposed to the results reported for systems which only involve the optomechanical coupling \cite{Vit08}.

\section{Conclusion}
We analyzed the squeezing and optomechanical entanglement in electromechanical system in which a superconducting charge qubit is coupled to a transmission line resonator and a movable membrane, which in turn is coupled to a second transmission line resonator. We show that due the nonlinearities induced by the optomechanical coupling and the superconducting qubit, the transmitted microwave field exhibits strong squeezing. Besides, we showed that robust optomechanical entanglement can be achieved by tuning the bath temperature of the two resonators. We also showed that the generated entanglement can be controlled for appropriate choice of the input drive pump power and the detuning of the drive frequency from the resonator frequency. Merging of optomechanics with electrical circuits opens new avenue for an alternative way to explore creation and manipulation of quantum states of microscopic systems.

\begin{acknowledgements}
The authors thank Konstantin Dorfman for useful discussions. This research was funded by the
Office of the Director of National Intelligence (ODNI), Intelligence
Advanced Research Projects Activity (IARPA), through the Army
Research Office Grant No. W911NF-10-1-0334. All statements of fact,
opinion or conclusions contained herein are those of the authors and
should not be construed as representing the official views or
policies of IARPA, the ODNI, or the U.S. Government. We also
acknowledge support from the ARO MURI Grant No. W911NF-11-1-0268.
\end{acknowledgements}

\appendix
\section{The terms that appear in Eq. \eqref{sda}}
The coefficients that appear in Eq. \eqref{sda} are given by
\begin{align}\label{a1}
  \xi_{1}&=\frac{\sqrt{\kappa_{a}}}{d(\omega)}\eta_{-}[(u_{-} v_{-} - |\mathcal{B}|^2) (u_{+} v_{+} - |\mathcal{B}|^2) -
   u_{-} u_{+} |\mathcal{C}|^2]\notag\\
  &-\frac{\sqrt{\kappa_{a}}|G|^2}{d(\omega)}\big\{[(u_{-} -u_{+}) |\mathcal{B}|^2 +
    (|\mathcal{B}|^2 \notag\\
    &+u_{-}u_{+} [v_{-}-v_{+} + 2i\text{Im}(\mathcal{C}) ]\big\},
\end{align}
\begin{align}\label{a2}
 \xi_{2}=&\frac{\sqrt{\kappa_{a}}G^2}{d(\omega)}\Big\{[(u_{-}-u_{+})|\mathcal{B}|^2\notag\\
 &+u_{-}u_{+} [v_{-}-v_{+} + 2i\text{Im}(\mathcal{C})]\Big\},
 \end{align}
 \begin{equation}\label{a3}
 \xi_{3}= \frac{\sqrt{\gamma_{\rm m}}G\eta_{-}}{d(\omega)}u_{+}\left(|\mathcal{B}|^2+u_{-}\mathcal{C}^{*}-u_{-}v_{-}\right),
 \end{equation}
 \begin{equation}\label{a4}
  \xi_{4}= \frac{\sqrt{\gamma_{\rm m}}G}{d(\omega)}\eta_{-}u_{-}\left(|\mathcal{B}|^2+u_{+}\mathcal{C}^{*}-u_{+}v_{+}\right),
 \end{equation}
  \begin{equation}\label{a5}
 \xi_{5}=-\frac{\sqrt{\kappa_{c}}G}{d(\omega)}\eta_{-}\mathcal{B}^{*}\left(|\mathcal{B}|^2+u_{-}\mathcal{C}^{*}-u_{-}v_{-}\right),
 \end{equation}
 \begin{equation}\label{a5}
 \xi_{6}=-\frac{\sqrt{\kappa_{c}}G}{d(\omega)}\eta_{-}\mathcal{B}\left(|\mathcal{B}|^2+u_{-}\mathcal{C}-u_{+}v_{+}\right),
 \end{equation}
where
\begin{align}
d(\omega)=&[(u_{-}v_{-}-|\mathcal{B}|^2)(v_{+}u_{+}-|\mathcal{B}|^2)-u_{-}u_{-}|\mathcal{C}|^2]\eta_{-}\eta_{+}\notag\\
&+|G|^2\Big\{u_{-}[|\mathcal{B}|^2+u_{+}(v_{-}-v_{+}+2i\text{Im}(\mathcal{C}))]\notag\\
&-u_{+}|\mathcal{B}|^2\Big\}(\eta_{-}-\eta_{+}).
\end{align}

\end{document}